\documentclass[aps,prl,twocolumn,superscriptaddress,showpacs,10pt]{revtex4-1}

\usepackage{bm}
\usepackage{graphicx}
\usepackage{longtable}
\usepackage{amsmath}
\usepackage{amssymb}
\usepackage{color}

\usepackage{hyperref}

\begin{document}

\title{Upper bound for the $s$--$d$ exchange integral in $n$-(Ga,Mn)N:Si from magnetotransport studies}

\author{R. Adhikari}
\affiliation{Institut f\"ur Halbleiter-und-Festk\"orperphysik, Johannes Kepler University, Altenbergerstr. 69, A-4040 Linz, Austria}

\author{W. Stefanowicz}
\affiliation{Institute of Physics, Polish Academy of Sciences, al. Lotnik\'ow 32/46, PL-02 668 Warszawa, Poland}

\author{B. Faina}
\affiliation{Institut f\"ur Halbleiter-und-Festk\"orperphysik, Johannes Kepler University, Altenbergerstr. 69, A-4040 Linz, Austria}

\author{M.~Sawicki}
\affiliation{Institute of Physics, Polish Academy of Sciences, al. Lotnik\'ow 32/46, PL-02 668 Warszawa, Poland}

\author{T. Dietl}
\email{dietl@ifpan.edu.pl}
\affiliation{Institute of Physics, Polish Academy of Sciences, al. Lotnik\'ow 32/46, PL-02 668 Warszawa, Poland}
\affiliation{Institute of Theoretical Physics, Faculty of Physics, University of Warsaw,
ul. Pasteura 5, PL-02-093 Warszawa, Poland}
\affiliation{WPI-Advanced Institute for Materials Research (WPI-AIMR), Tohoku University, 2-1-1 Katahira, Aoba-ku, Sendai 980-8577, Japan}

\author{A. Bonanni}
\email{alberta.bonanni@jku.at}
\affiliation{Institut f\"ur Halbleiter-und-Festk\"orperphysik, Johannes Kepler University, Altenbergerstr. 69, A-4040 Linz, Austria}

\begin{abstract}
A series of recent magnetooptical studies pointed to contradicting values of the $s$--$d$ exchange energy $N_0\alpha$ in Mn-doped GaAs and GaN as well as in Fe-doped GaN. Here, a strong sensitivity of weak-localization phenomena to symmetry breaking perturbations (such as spin-splitting and spin-disorder scattering) is exploited to evaluate the magnitude of $N_0\alpha$ for $n$-type wurtzite (Ga,Mn)N:Si films grown by metalorganic vapor phase epitaxy. Millikelvin magnetoresistance studies and their quantitative interpretation point to $N_0\alpha < 40$\,meV, a value at least 5 times smaller than the one found with similar measurements on, e.g., n-(Zn,Mn)O.  It is shown that this striking difference in the values of the $s$--$d$ coupling between $n$-type III-V and II-VI dilute magnetic semiconductors can be explained by a theory that takes into account the acceptor character of Mn in III-V compounds.
\end{abstract}

\date{\today}

%75.50.Pp Magnetic semiconductors
%71.70.Ej Spin-orbit coupling, Zeeman and Stark splitting, Jahn-Teller effect
%72.15.Rn Localization effects (Anderson or weak localization)
%72.80.Ey III-V and II-VI semiconductors
%72.20.-i Conductivity phenomena in semiconductors and insulators

\maketitle

\section{Introduction}

The existence of a strong $sp$--$d$ interaction $-J\vec{s}\cdot\vec{S_i}$ between effective mass carriers with a spin $\vec{s}$ and a subsystem of spins $\vec{S}_i$ localized on magnetic impurities is characteristic of dilute magnetic semiconductors (DMSs) \cite{Furdyna:1988_B,Dietl:1994_B,Gaj:2010_B}.  This interaction accounts for the giant Zeeman
splitting of bands, spin-disorder scattering, the formation of magnetic polarons, and the mediation by itinerant carriers of ferromagnetic coupling between magnetic ions \cite{Furdyna:1988_B,Dietl:1994_B,Gaj:2010_B,Dietl:2014_RMP}.  A strong sensitivity of quantum-localization phenomena to symmetry breaking perturbations (such as spin-splitting and spin-disorder scattering) results in striking magnetotransport phenomena in DMSs \cite{Dietl:2008_JPSJ}. In this work we exploit this sensitivity and demonstrate that the $s$--$d$ exchange integral is surprisingly small in Ga$_{1-x}$Mn$_x$N:Si with low Mn content $x$. We explain this result in terms of a previously developed theoretical model \cite{Sliwa:2008_PRB} that links the reduction in the apparent $s$--$d$ interaction to the acceptor character of Mn in GaN.

In the case of tetrahedrally bound semiconductors, the $sp$--$d$ interaction is characterized by the two exchange integrals,
$\alpha = \langle S|J|S\rangle$ and $\beta = \langle X|J|X\rangle$, where $S$ and $X$ stand for the periodic parts of the Bloch wave function (Kohn-Luttinger amplitudes), which respectively transform as  atomic $s$ and $p_x$ orbitals under symmetry operations of the crystallographic point group \cite{Dietl:1994_B,Kacman:2001_SST}.

If the bottom of the conduction band is formed by states derived from cation and anion $s$ orbitals, spin-dependent effects involving electrons originate from the intra-atomic potential $s$--$d$ exchange interaction. Indeed, the corresponding $s$--$d$ exchange energy is 392\,meV in the case of free Mn$^{1+}$ ions \cite{Dietl:1994_PRB}, whereas
in DMSs this value is subjected to an up to twofold reduction by a covalent admixture of the anion $s$-type wave function to the Kohn-Luttinger amplitude at the conduction band edge.
 In accord with this insight, the values of $N_0\alpha$ (where $N_0$ is the cation concentration) as determined from magnetooptics experiments, are in the range $190 \lesssim N_0\alpha \lesssim 320$\,meV for a series of paramagnetic II-VI compounds doped with Mn as well as with Cr, Fe, and Co \cite{Kacman:2001_SST,Whitaker:2011_NL}. These findings are in good quantitative agreement with {\em ab initio} studies for (II,Mn)VI DMSs \cite{Kim:2006_JKPS,Chanier:2009_PRB}. At the same time, by employing the magnetooptical values of $N_0\alpha$ it was possible to quantitatively interpret the positive magnetoresistance originating from the influence of giant spin-splitting of the conduction band upon disorder-modified electron-electron interactions in paramagnetic $n$-type (Cd,Mn)Se \cite{Sawicki:1986_PRL}, (Cd,Mn)Te \cite{Shapira:1990_PRB,Jaroszynski:2007_PRB}, (Cd,Zn,Mn)Se \cite{Smorchkova:1997_PRL}, (Zn,Mn)O \cite{Andrearczyk:2005_PRB}, and (Zn,Co)O \cite{Dietl:2007_PRB} near the metal-to-insulator transition.

In contrast to this clear cut situation, the physics of exchange coupling between conduction band electrons and magnetic impurities in III-V DMSs appears more complex. Here, {\em ab initio} computations predicted that the magnitudes of $N_0\alpha$ in (Ga,Mn)As and (Ga,Mn)N should be similar to those in (II,Mn)VI DMSs \cite{Kim:2006_JKPS,Dalpian:2006_PRB}. However, rather different values are implied by magnetooptical studies on dilute paramagnetic Ga$_{1-x}$Mn$_x$As ($x \leq 0.13$\%) grown by molecular beam epitaxy at high temperatures, which point to {\em antiferromagnetic} $N_0\alpha   = (-23 \pm 8$)\,meV for photoelectrons at the band-edge \cite{Myers:2005_PRL,Poggio:2005_PRB,Stern:2007_PRB}.
These data challenge the time-honored notion that the spin-dependent coupling between the electrons and Mn spins in a tetrahedrally coordinated DMS originates from the necessarily ferromagnetic intra-atomic potential $s$--$d$ exchange. This riddle was solved \cite{Sliwa:2008_PRB} by noting that for the Mn concentrations and growth conditions in questions, Mn ions are accompanied by bound holes. In such a case, in addition to  the $s$--$d$ interaction,   $s$--$p$ exchange between electrons and bound holes is present and overcompensates the $s$--$d$ coupling. Good agreement between experimental and theoretical values of $N_0\alpha$ was found without adjustable parameters \cite{Sliwa:2008_PRB}.
This model explained also (i) the two orders of magnitude {\em longer} spin relaxation time of electrons in GaAs:Mn comparing to GaAs:Ge, where only $s$--$p$ exchange operates \cite{Astakhov:2008_PRL}; (ii) the crossover to a positive value $xN_0\alpha = +2.3\times 10^{-2}$\,meV in presumably more compensated Ga$_{1-x}$Mn$_x$As samples ($x \simeq 0.1$\%),  in which electron spin-flip Raman scattering was observed \cite{Heimbrodt:2001_PE}, and (iii) a much reduced spin splitting  found for electrons injected to InAs quantum dots containing a neutral Mn acceptor \cite{Kudelski:2007_PRL}.

In the case of magnetically doped wurtzite (wz) GaN, exciton magnetospectroscopy \cite{Pacuski:2007_PRB,Pacuski:2008_PRL,Suffczynski:2011_PRB}
and magnetic circular dichroism \cite{Rousset:2013_PRB}  were employed to evaluate the $sp$--$d$ exchange energies in films deposited by metal-organic vapor phase epitaxy (MOVPE). These methods provide the magnitude of $N_0(\beta-\alpha)$ rather accurately. If, however, all three fundamental excitons A, B, and C are resolved some information on $N_0\alpha$ can also be obtained.
By applying a multiparameter fitting procedure to magnetoreflectivity spectra the values $N_0\alpha =  (0 \pm 100)$ and $(+100 \pm 200)$\,meV were determined for Ga$_{1-x}$Fe$_x$N ($x\leq 0.21$\%) and Ga$_{1-x}$Mn$_x$N ($x\leq 0.9$\%), respectively \cite{Suffczynski:2011_PRB,Pacuski:2008_PRL}.
The former was reassessed by examining excitonic magnetic circular dichroism for Ga$_{1-x}$Fe$_x$N with $x=0.2$\%, which resulted in $N_0\alpha = (+50 \pm 100)$\,meV \cite{Rousset:2013_PRB}.
In contrast, results of more recent time-resolved Kerr rotation measurements that probe directly photoelectrons, not excitons, point to a standard value $N_0\alpha = (+230 \pm 20)$\,meV for a series of Ga$_{1-x}$Mn$_x$N films obtained by MOVPE with $x$ up to $0.27$\% \cite{Hsu:2014_PRB}. On the other hand, a much lower value $N_0|\alpha| = (14 \pm 4)$~meV was found by  analyzing  the effect of the electrons on the Mn$^{2+}$ longitudinal relaxation time $T_1$ in compensated bulk $n$-type Ga$_{1-x}$Mn$_x$N samples with $ x \le 0.2$\% \cite{Wolos:2003_APL}. However, the interpretation
of electron paramagnetic resonance data was carried out \cite{Wolos:2003_APL} neglecting possible effects of the relaxation-time bottleneck \cite{Barnes:1981_AP} that in the case of similar Land\'e factors of carriers and localized spins can increase the apparent value of $T_1$, leading to an underestimation of the $s$--$d$ coupling energy $N_0|\alpha|$.

The above discussion indicates that the issue of the $s$--$d$ interaction in (Ga,Mn)N, and more generally in III-V DMSs, is by no means settled. Here we present results of millikelvin magnetoconductance (MC) measurements on a series of $n$-type wz-(Ga,Mn)N:Si films grown by MOVPE. The studied samples have nominally the same concentration of Si donors, $N_D \approx 10^{19}$\,cm$^{-3}$, but different content $x$ of Mn acceptors, so that the electron concentration diminishes with $x$ down to $n =1.4\times 10^{18}$ for $x= 0.06$\%. The MC data are described in terms of a model that takes quantum-localization corrections into account \cite{Altshuler:1985_B,Fukuyama:1985_B,Lee:1985_RMP,Ono:1983_JPSJ}, and which previously described successfully MC experiments on $n$-type II-VI DMSs \cite{Sawicki:1986_PRL,Shapira:1990_PRB,Jaroszynski:2007_PRB,Smorchkova:1997_PRL,Andrearczyk:2005_PRB,Dietl:2007_PRB} as well as on their non-magnetic counterparts, i.e., zinc-blende CdTe:In \cite{Jaroszynski:1997_TSF} and wurtzite CdSe:In \cite{Sawicki:1986_PRL}, ZnO:Al \cite{Andrearczyk:2005_PRB}, and GaN:Si, as we reported recently \cite{Stefanowicz:2014_PRB}. In particular, we take here into account the influence of spin-splitting of the conduction band upon disorder-induced one-electron \cite{Fukuyama:1985_B,Ono:1983_JPSJ} and many-body interference phenomena \cite{Altshuler:1985_B,Fukuyama:1985_B,Lee:1985_RMP}. Our present results  indicate that the MC of Ga$_{1-x}$Mn$_x$N:Si is not markedly affected by $s$--$d$ coupling. Their quantitative interpretation points to $N_0|\alpha| < 40$\,meV. Thus our data provide a new support to the theory \cite{Sliwa:2008_PRB} that assigns a reduction in the apparent magnitude of $N_0\alpha$ to repulsion of electrons by negatively charged Mn acceptors in $n$-type III-V DMSs.

\section{Samples and experiment}

The GaN layers co-doped with Si and Mn have been grown in an AIXTRON 200RF horizontal tube metalorganic vapor phase epitaxy (MOVPE) reactor and deposited on a $c$-plane sapphire substrate using TMGa, MnCp$_{2}$, NH$_{3}$, and SiH$_{4}$ as precursors for Ga, Mn, N, and Si respectively, with H$_{2}$ as carrier gas. After nitridation of the sapphire substrate, a low temperature nucleation layer (NL) is deposited at 540$^{\circ}$C and then annealed at 1040$^{\circ}$C. An insulating 1\,$\mathrm{\mu}$m-thick GaN:Mn buffer layer is grown at 1040$^{\circ}$C, Mn being introduced in order to compensate residual donors accounting for the $n$-type character of conductivity in GaN fabricated by MOVPE. Finally, a (Ga,Mn)N:Si layer with a thickness $d=150$\,nm is grown at $1000^{\circ}$C.  The Mn content $x$ ranges from about 0.01 to 0.06\%, as determined by secondary ions mass spectroscopy (SIMS).

The grown samples are systematically characterized by atomic force microscope (AFM), high-resolution x-ray diffraction (HRXRD), and high-resolution transmission electron microscopy (HRTEM). The AFM micrographs reveal a flat surface (rms roughness $\approx$ 1\,nm) while HRXRD and HRTEM confirm the high crystallinity of the samples.
The HRTEM analysis does not reveal any secondary phases like, e.g., precipitates of Si$_\mathrm{x}$N or Mn-rich compounds.

Hall bars with Ti/Au/Al/Ti/Au metallic contacts have been fabricated by conventional photolithography. Measurements of resistivity  $\rho(T,B)$ as a function of temperature $T$ and magnetic field $B$ have been performed in a dedicated home-built helium cryostat and a dilution refrigerator down to 50\,mK. To avoid a remanent magnetic field characteristic to superconducting magnets, weak field measurements are carried out in a copper coil.

The magnitudes of electron concentration $n$, mobility $\mu$, and the product of the Fermi wave vector and the mean free path, $k_{\text{F}}\ell$, obtained from Hall resistivity measurements at 5\,K, and SIMS's determined $x$ are collected in Table\,\ref{tab:tab1} for the sample S0 discussed peviously \cite{Stefanowicz:2014_PRB} and for the Mn-doped GaN:Si samples S1-S3 studied within the present work. For comparison, $x$ values estimated from the Hall data according to $x(\text{Hall}) = [n(\text{S0}) -n]/N_0$ are also shown in Table\,\ref{tab:tab1}. It is assumed in this evaluation that the incorporation of Si donors is independent of the Mn acceptor density.

As seen, by co-doping with Mn acceptors we drive GaN:Si towards the metal-to-insulator transition occurring at $n \simeq 1\times 10^{18}$\,cm$^{-3}$ in n-GaN \cite{daSilva:2002_JAP}. In this range, the magnitude of conductance is strongly affected by quantum-localization corrections that depend sensitively, according to theoretical expectations \cite{Altshuler:1985_B,Lee:1985_RMP,Fukuyama:1981_JPSJ,Ono:1983_JPSJ} and experimental studies \cite{Sawicki:1986_PRL,Shapira:1990_PRB,Jaroszynski:2007_PRB,Smorchkova:1997_PRL,Andrearczyk:2005_PRB,Dietl:2007_PRB}, on the giant spin-splitting of the conduction band in DMSs.

%%%%%%%%%%%%%%%%%%%%%%%%%%%%%%%%%%%%%%%%%%
\begin{table}[tb]
	\caption{Parameters of the reference sample S0 (Ref.\,\onlinecite{Stefanowicz:2014_PRB}) and
of the samples investigated in this work S1--S3.}
\begin{tabular}{c c c c c c c}
		\hline \hline
     Sample  &        $n$         &     $\mu$     &   $k_{\text{F}}\ell$ &$x$(\%)&     $x$(\%)     \\
                  &  ($10^{18}$ cm$^{-3}$)   &   (cm$^{2}/$Vs)   &             &  SIMS    &   Hall \\ \hline
    S0  &  12    &  140  &   4.6        &  $<0.01$                & refr.    \\
    S1  &  7.3   &  139  &   3.3        &  $\lesssim 0.01$     &  0.011         \\
    S2  &  4.7   &  103  &   1.8        &  0.02                      &   0.017       \\
    S3  &  1.4   &  102  &   0.8      &  0.06                      &   0.024      \\  \hline \hline
	\end{tabular}
	\label{tab:tab1}
\end{table}
%%%%%%%%%%%%%%%%%%%%%%%%%%%%%%%%%%%%%%%%%%

\section{Results }

\begin{figure}[bt]
	\centering
	\includegraphics[width=0.8\linewidth]{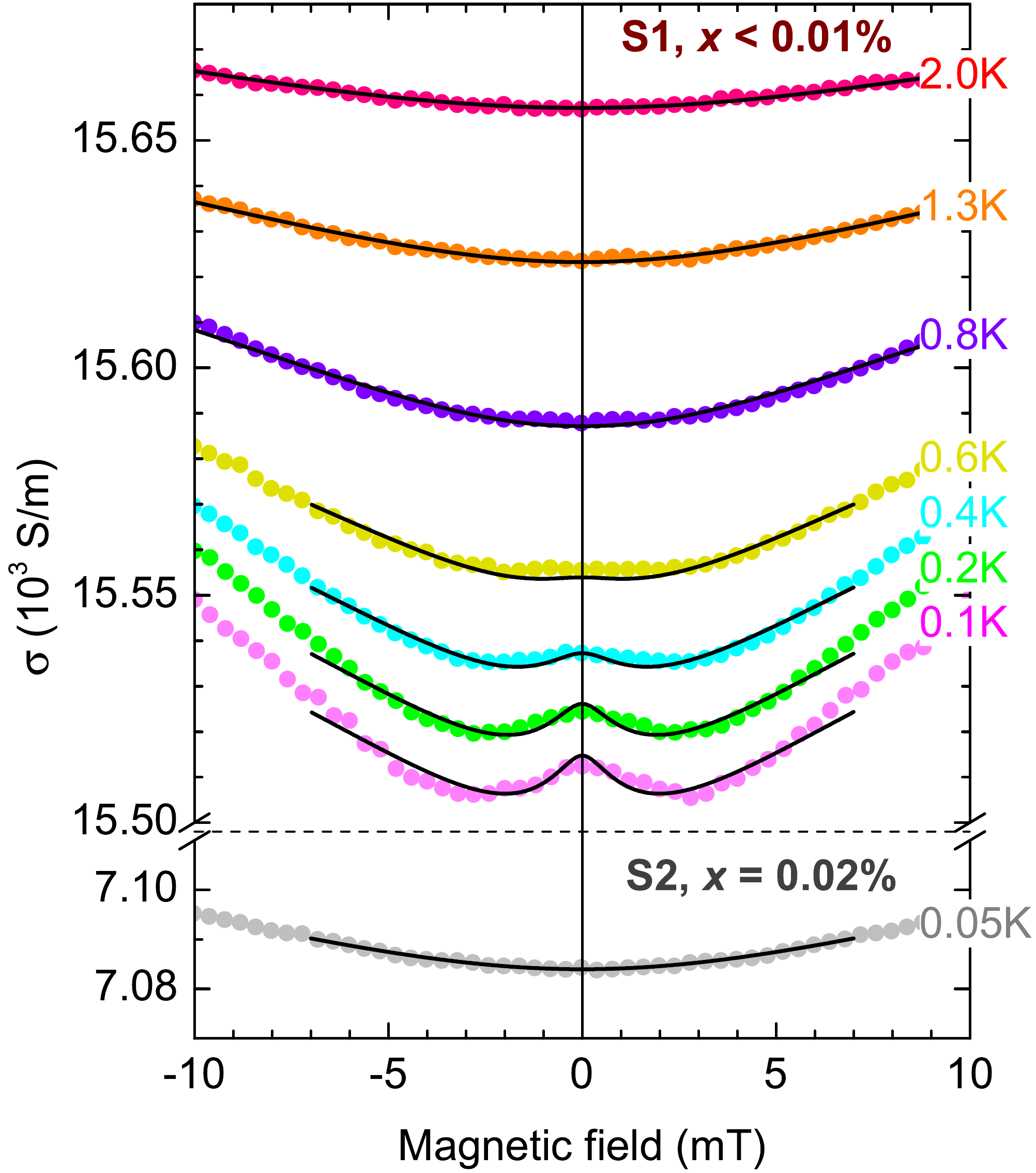}
	\caption{(Color online) Dots: measured magnetoconductivity for sample S1 at $T \leq 2$\,K  and for sample S2 at $T=0.05$\,K. Solid black lines show results of  fitting within the theoretical 2D model treating $L_{\varphi}(T)$ as the only fitting parameter.}
	\label{fig:fig1}	
\end{figure}

The magnitude of the conductivity $\sigma(T,B) =1/\rho(T,B)$ has been measured at various temperatures as a function of the magnetic field applied perpendicular to the film surface (i.e., parallel to the wz-$c$-axis). As seen in Fig.\,\ref{fig:fig1}  both negative and positive MC is visible in low magnetic fields below 0.6\,K for the sample S1 with the lowest Mn content, i.e., with the highest electron concentration. The negative component of MC is related to the onset of a weak antilocalization maximum, a distinct signature of spin-orbit coupling. For the samples S2 and S3 only a positive contribution to the MC is observed.

%%%%%%%%%%%%%%%%%%%%%%%%%%%%%%%%%%%%%%%%%%

\begin{figure}[bt]
	\centering
	\includegraphics[width=0.8\linewidth]{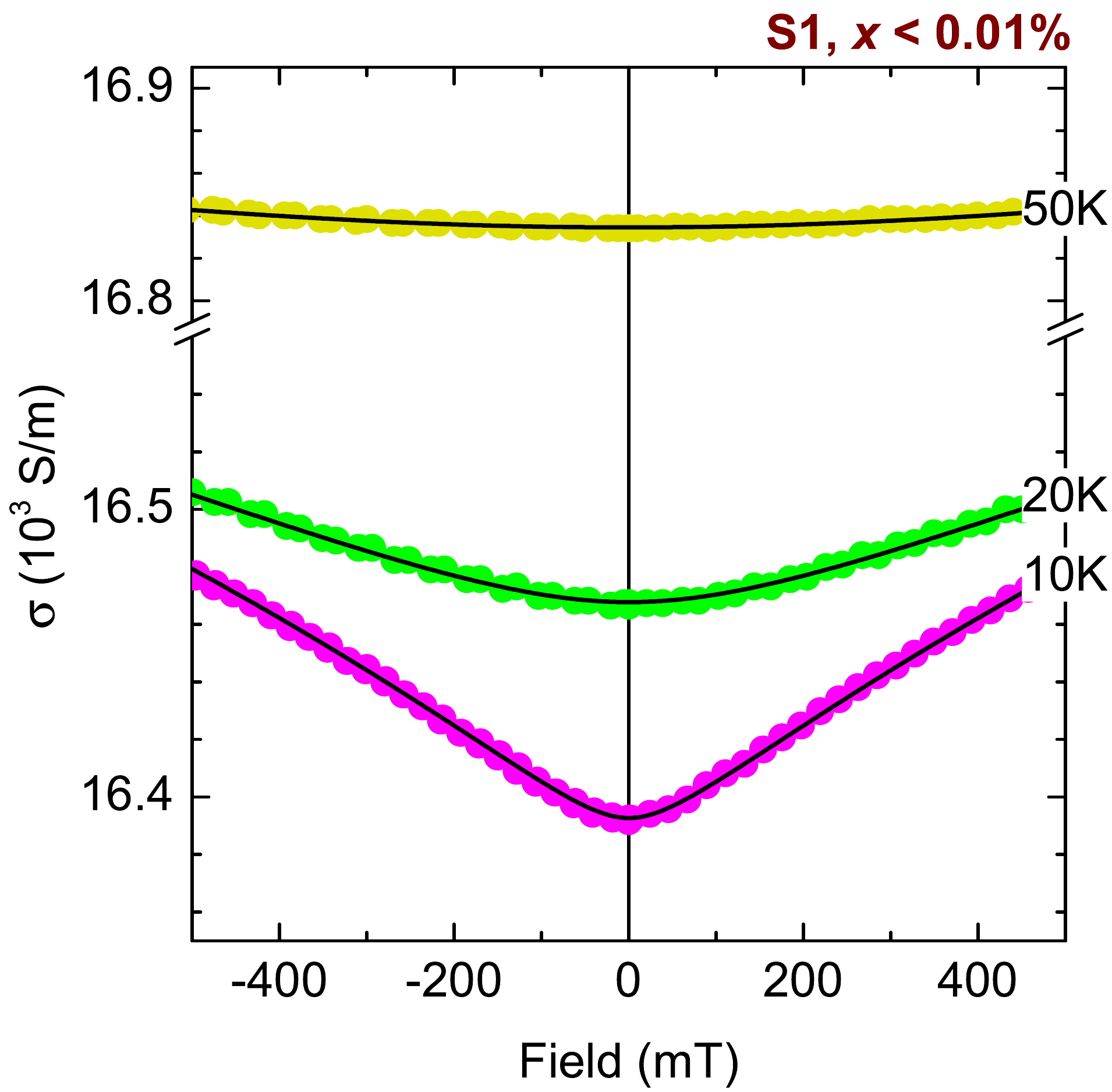}
	\caption{(Color online) Magnetoconductivity measured for sample S1 at $T \geq 10$\,K. Bullets: experimental data; solid lines: theoretical fitting within the 3D theory with the phase coherence length $L_{\varphi}(T)$ as the only fitting parameter.}
	\label{fig:fig2}	
\end{figure}
%%%%%%%%%%%%%%%%%%%%%%%%%%%%%%%%%%%%%%%%%%

The data are analyzed by the two and three dimensional (2D and 3D, respectively) approaches developed for MC in the weakly localized regime, $k_F \ell>1$, as the dimensional crossover condition for one-electron interference phenomena is expected to occur at relevant temperatures in our thin films. In terms of the phase coherence length $L_{\varphi} = (D\tau_{\varphi})^{1/2}$, where  $D=\hbar k_F\ell/3m^*$ is the the diffusion coefficient, and   $\tau_{\varphi}$ is the phase coherence time, the crossover occurs at $L_{\varphi}  \simeq d$, where the film thickness $d = 150$\,nm in our films. At the same time we neglect a possible crossover in the case of electron-electron interaction effects, as it occurs at lower temperatures, $L_T  \simeq d$, where $L_T = (\hbar D/k_BT)^{1/2}$. We use the value of the Rashba spin-orbit coupling constant $\alpha_R=4.5$\,meV{\AA} we determined previously for GaN:Si \cite{Stefanowicz:2014_PRB}.

%%%%%%%%%%%%%%%%%%%%%%%%%%%%%%%%%%%%%%%%%%
\begin{figure}[bt]
	\centering
	\includegraphics[width=1.0\linewidth]{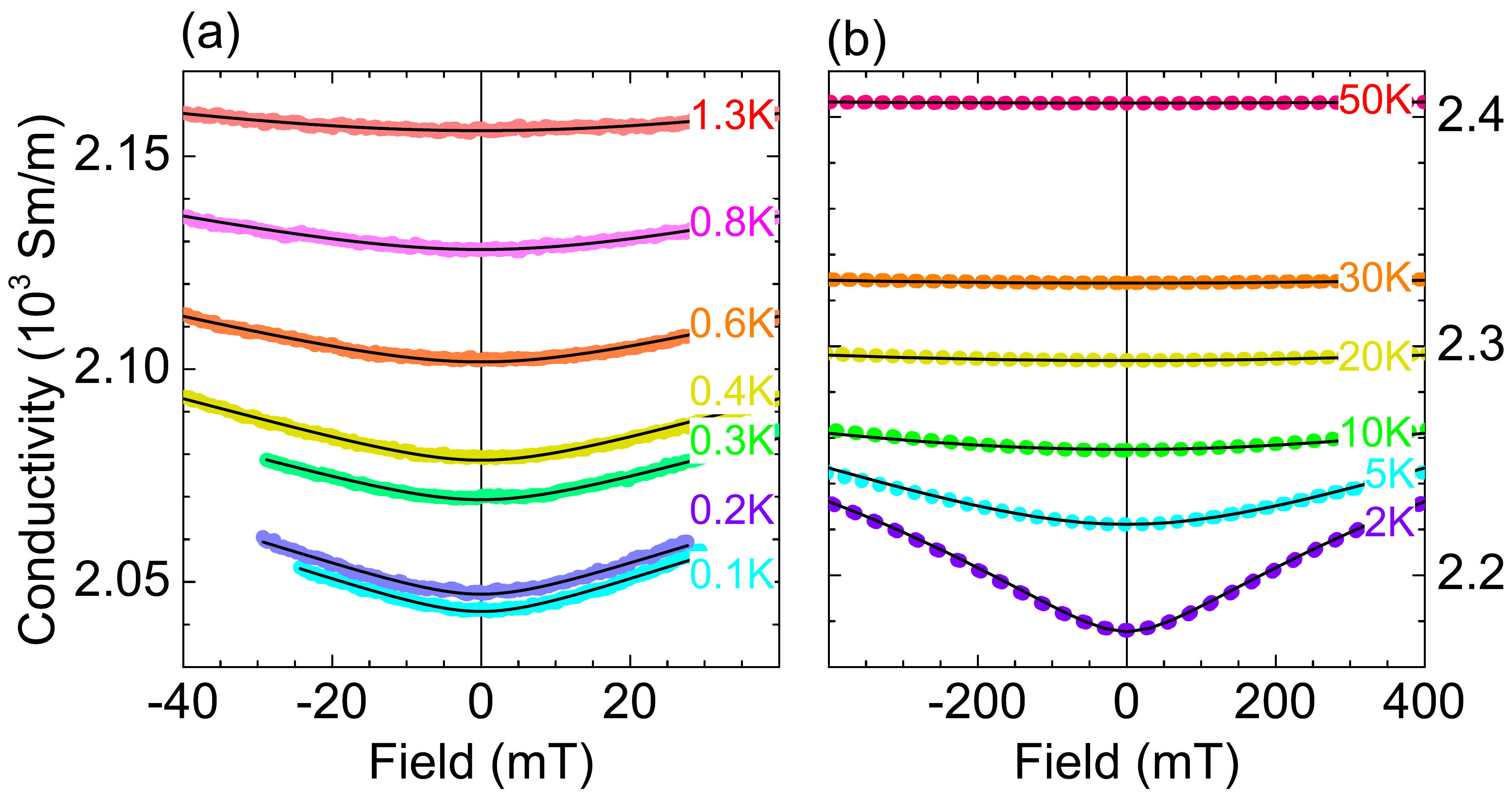}
	\caption{(Color online) Magnetoconductivity measured for sample S3 at various temperatures. Dots: experimental data; solid lines: theoretical fitting within the 3D theory with the phase coherence length $L_{\varphi}(T)$ as the only fitting parameter.}
	\label{fig:fig3}	
\end{figure}
%%%%%%%%%%%%%%%%%%%%%%%%%%%%%%%%%%%%%%%%%%

At first we disregard the presence of the $s$--$d$ interaction, i.e., we neglect both the exchange contribution to spin-splitting and spin-disorder scattering. Therefore, $L_{\varphi}(T)$ is the only fitting parameter. According to the data presented in Figs.\,\ref{fig:fig1}--\ref{fig:fig3}, the 2D or the 3D model employed in the respective temperature ranges, describe the experimental data very well. 
We have checked that if the spin-splitting is given by $\hbar\omega_s = g^*\mu_BB$, where $g^* = 1.95$ in GaN \cite{Carlos:1993_PRB}, one-electron localization phenomena dominate, and effects of spin-splitting upon one-electron and many-electron contributions are irrelevant to the MC.

\begin{figure}[bt]
	\centering
	\includegraphics[width=0.8\linewidth]{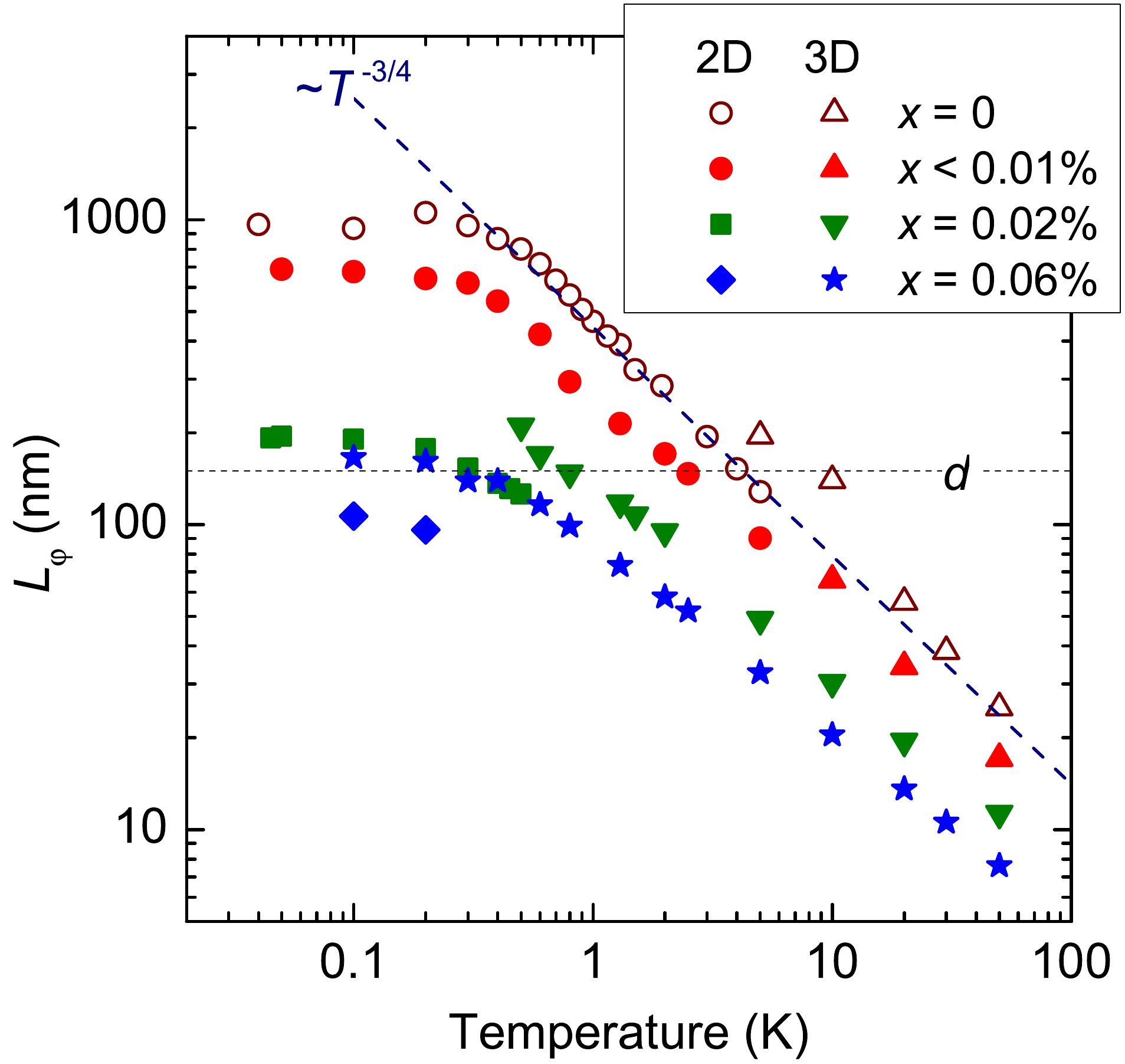}
	\caption{(Color online) Temperature dependence of phase coherence length $L_{\varphi}$ obtained by fitting magnetoconductance data for (Ga,Mn)N:Si layers within the 2D and 3D models. For comparison the corresponding data for \emph{n}-GaN:Si \cite{Stefanowicz:2014_PRB} is also presented (open symbols). Dashed line: $T^{-3/4}$ dependence. The horizontal line marks the layer thickness $d=150$\,nm.}
	\label{fig:fig4}	
\end{figure}

The fitted values of $L_{\varphi}(T)$ are summarized in Fig.\,\ref{fig:fig4}. The data on  $L_{\varphi}(T)$ allow us to evaluate the temperature value corresponding to the dimensional crossover for particular samples, i.e., to assess the temperature ranges at which either the 2D or the 3D theory should be applied. Actually, this information has been exploited to select the appropriate model to fit the data collected in Figs.\,\ref{fig:fig1} -- \ref{fig:fig3}. Furthermore, as expected for decoherence associated with electron-electron interactions \cite{Altshuler:1985_B}, we find $L_{\varphi} = aT^{3/4}$, where $a$ increases with $k_F\ell$ \cite{Stefanowicz:2014_PRB}. This increase in decoherence with the approaching of the metal-insulator transition makes the effect of weak antilocalization invisible for the samples S2 and S3. Moreover, a change in the slope of $L_{\varphi}(T)$ is observed at about 0.3\,K, which, if not caused by noise-related decoherence, may signalize the dimensional crossover for effects associated with electron-electron interactions,  $L_T \simeq d$.  Finally, the values and the temperature dependence of $L_{\varphi}$ indicate that spin-disorder scattering \cite{Hikami:1980_PTP,Sawicki:1986_PRL}--for which the relaxation time is independent of temperature for paramagnetic spins--is of minor importance in the case treated here.

%%%%%%%%%%%%%%%%%%%%%%%%%%%%%%%%%%%%%%%%%%
%%%%%%%%%%%%%%%%%%%%%%%%%%%%%%%%%%%%%%%%%%

%%%%%%%%%%%%%%%%%%%%%%%%%%%%%%%%%%%%%%%%%%
\begin{figure}[bt]
	\centering
	\includegraphics[width=1.0\linewidth]{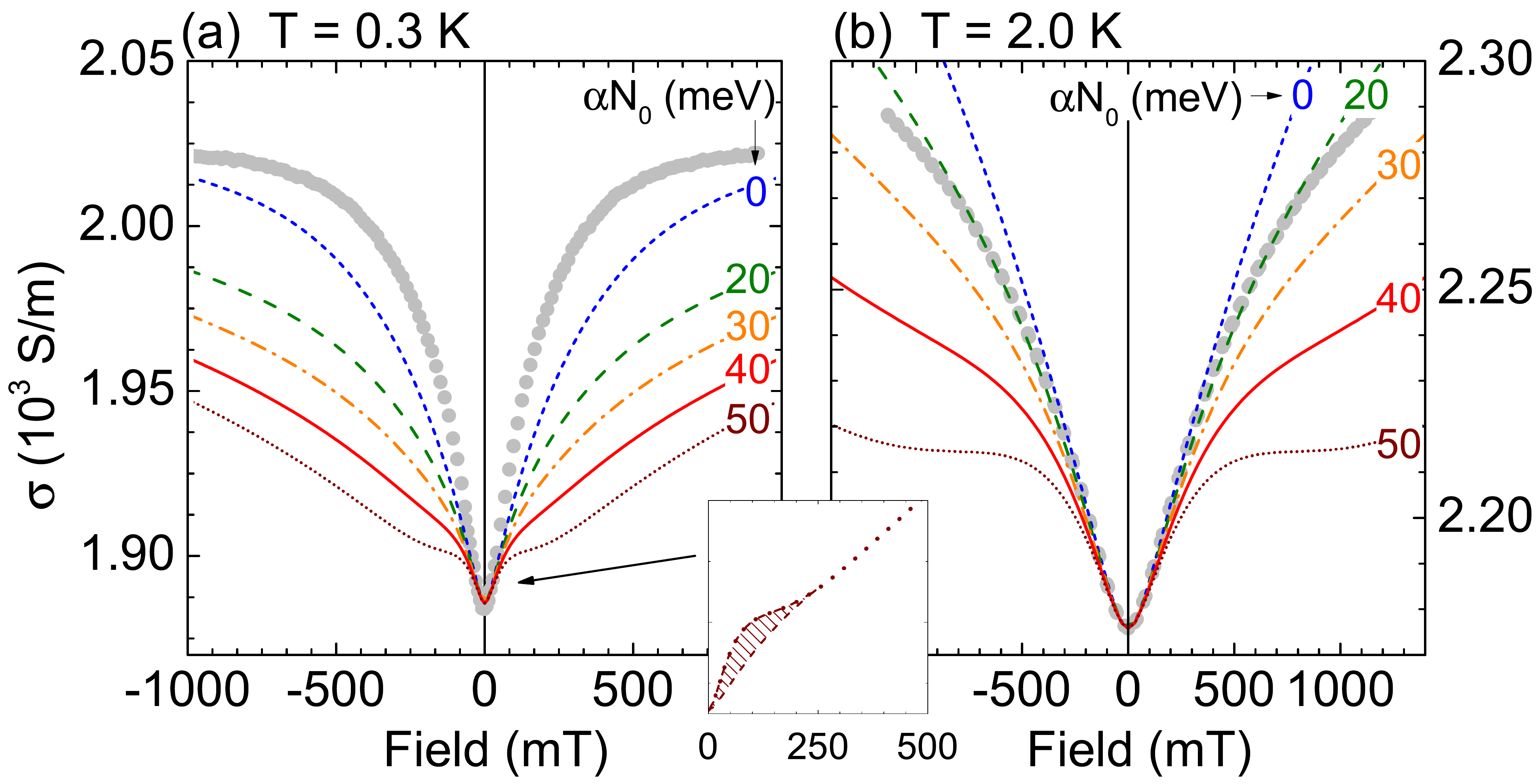}
	\caption{(Color online) Magnetoconductivity (dots) measured for sample S3 at 0.3\,K (a) and at 2\,K (b). Lines show values calculated for different magnitudes of $N_0\alpha $ taking the presence of Mn spins with concentration $x = 0.06$\% into account. For $N_0\alpha \gtrsim 40$\,meV a positive shoulder of magnetoconductivity (shown in the inset as a hatched area)  appears in the calculations.}
	\label{fig:fig5}
\end{figure}
%%%%%%%%%%%%%%%%%%%%%%%%%%%%%%%%%%%%%%%%%%

In order to estimate an upper limit of $N_0\alpha $  we compute MC for (Ga,Mn)N:Si taking the presence of $s$--$d$ exchange interaction into account. We employ the 3D model \cite{Altshuler:1985_B,Sawicki:1986_PRL}, and calculate the magnitude of MC for different values of $N_0\alpha$. In Fig.\,\ref{fig:fig5} the results for $T=0.3$\,K and 2\,K areshown in a relatively wide field range in comparison with the experimental data for the sample S3 with the highest Mn concentration $x =0.06$\%.   The emergence of a significant negative contribution to MC, and of a kink, are seen in the simulations performed for $N_0\alpha  \gtrsim 20$ and 40\,meV, respectively. The absence of such features in the experimental results indicates that in (Ga,Mn)N:Si the value of  $N_0|\alpha|$ is much smaller than $40$\,meV. Since the Mn effect scales with $xN_0|\alpha|$, in light of the data in Fig.\,\ref{fig:fig5}, the conclusion that $N_0|\alpha|  < 40$\,meV remains valid even if we assume $x=0.024$\%, as implied by the Hall data. Hence, our experimental results for n-(Ga,Mn)N corroborate the theoretical predictions for Mn-based III-V DMSs co-doped with shallow donors \cite{Sliwa:2008_PRB}. We also note that the spin-disorder scattering rate, $\tau_{s}^{-1}$ evaluated using the standard approach \cite{Hikami:1980_PTP,Sawicki:1986_PRL} is of the order of $10^6$\,s$^{-1}$, a value much smaller than the $\tau_{\varphi}^{-1} \sim 10^9$\,s$^{-1}$ of our samples. However, this conclusion is valid as long as spin-disorder scattering by magnetization fluctuations associated with the carrier density fluctuations in the vicinity of the MIT is not relevant \cite{Dietl:2008_JPSJ}. Due to the low Mn concentrations in our samples, the last contribution to spin-disorder scattering rate is expected to play a minor role.

\section{Discussion and conclusions}

We have carried out low temperature magnetotransport studies on high quality $n$-type wz-(Ga,Mn)N:Si films grown on a semi-insulating GaN:Mn buffer layer.  The quantitative model of  magnetoconductance in the weakly localized regime allows us to evaluate an upper limit for the magnitude of the $s$--$d$ exchange energy $N_0|\alpha| < 40$\,meV.  This result substantiates the theoretical model \cite{Sliwa:2008_PRB} that assigns a reduction of the $s$--$d$ exchange energy in $n$-type III-V DMSs to the acceptor character of Mn in these systems. According to this model, when Mn impurities are negatively charged, they repel the conduction band electrons, with the effect of reducing the apparent magnitude of $s$--$d$ coupling. Surprisingly, small magnitudes of $N_0\alpha$ were also found in studies of magnetoexcitons in semiinsulating, not $n$-type, (Ga,Mn)N \cite{Suffczynski:2011_PRB} and (Ga,Fe)N \cite{Rousset:2013_PRB}. We explain these observations with the occupation of mid-gap states of transition metal ions by trapped photoelectrons under steady state illumination, which---similarly to the case of $n$-type (Ga,Mn)N---diminishes the $s$--$d$ splitting of the conduction band. In accord with this interpretation, time-resolved Kerr rotation measurements that probe the magnetic moments of conduction band electrons at times shorter than their lifetime, led to the standard value of the exchange energy, $N_0\alpha = (230 \pm 20)$\,meV \cite{Hsu:2014_PRB}.

\section*{Acknowledgments}

This work was supported by the Austrian Science Foundation --
FWF (P22477 and P24471),
by the National Science Centre (Poland) by OPUS grant 2013/09/B/ST3/04175,
and by the EU  $7^{\mathrm{th}}$  Framework Programmes:
CAPACITIES project REGPOT-CT-2013-316014 (EAgLE), and
FunDMS Advanced Grant of the European Research Council (Grant No.\,227690).

%\bibliographystyle{apsrev}
%\bibliography{alphaN0_GaMnN_PRB}


\end{document}